\documentclass[pr,twocolumn,nobibnotes,nofootinbib,notitlepage,superscriptaddress]{revtex4}
\usepackage{amsmath}
\usepackage{amssymb}
\usepackage{bm}
\usepackage{epsfig}
\usepackage{graphicx}
\usepackage [cp1251]{inputenc}
\usepackage[T2A]{fontenc}
\usepackage [russian,english]{babel}
\usepackage{color}
\usepackage[colorlinks]{hyperref}

\renewcommand{\Re}{\mathop{\rm Re}}

\begin{document}

\newcount\timehh  \newcount\timemm
\timehh=\time \divide\timehh by 60
\timemm=\time
\count255=\timehh\multiply\count255 by -60 \advance\timemm by \count255

\title{Linear optics, Raman scattering,  and spin noise spectroscopy}

\author{M. M. Glazov}
\affiliation{Ioffe Institute, 26 Polytechnicheskaya, St.-Petersburg 194021, Russia}
\affiliation{Spin Optics Laboratory, St. Petersburg State University, 1 Ul'anovskaya,
Peterhof, St. Petersburg 198504, Russia}

\author{V. S. Zapasskii}
\affiliation{Spin Optics Laboratory, St. Petersburg State University, 1 Ul'anovskaya,
Peterhof, St. Petersburg 198504, Russia}



\begin{abstract}
Spin noise spectroscopy (SNS) is a new method for studying magnetic resonance and spin dynamics based on measuring the Faraday rotation noise. In strong contrast with methods of nonlinear optics, the spectroscopy of spin noise is considered to be essentially nonperturbative. Presently, however, it became clear that the SNS, as an optical technique, demonstrates properties lying far beyond the bounds of conventional linear optics. Specifically, the SNS shows dependence of the signal on the light power density, makes it possible  to penetrate inside an inhomogeneously broadened absorption band and to determine its homogeneous width, allows one to realize an effective pump-probe spectroscopy without any optical nonlinearity, etc. This may seem especially puzzling when taken into account that SNS can be considered just as a version of Raman spectroscopy, which is known to be deprived of such abilities. In this paper, we clarify this apparent inconsistency.
\end{abstract}

\maketitle

\section{Introduction}

Potentialities of linear optics of material media are known to be determined by and to be restricted to a linear relationship between the electric field $\bm E$ of the light wave and the polarization $\bm P$ of the medium~\cite{ll8_eng} 
\begin{equation}
\label{PElin}
P_i = \chi_{ij} E_j.
\end{equation}
Here, $i,j=x,y,z$ are Cartesian components, $\chi_{ij} \equiv \chi_{ij}(\omega)$ is a  tensor complex quantity describing susceptibility of the medium, $\omega$ is the frequency of the incident radiation, and the polarization $\bm P$ in the linear approach is generated at the same frequency. Spectral dependence of the susceptibility $\chi_{ij}(\omega)$ provides the basis of the linear optical spectroscopy -- a highly powerful tool for studying electronic and atomic structure of materials as well as dynamics of charge carriers. A characteristic feature of the linear polarization $\bm P$ induced by the light field $\bm E$ is that it obeys the superposition principle, which implies, in particular, that the result of action of two fields $\bm E_1$ and $\bm E_2$ is identical to the sum of results of action of each of them separately:
\begin{equation}
\label{superposition}
\bm P(\bm E_1 + \bm E_2) = \bm P(\bm E_1) + \bm P(\bm E_2).
\end{equation}
In terms of measurable quantities, dependence~\eqref{PElin} is usually reduced to a linear relationship between the light \emph{intensity} and response of the medium with the intensity-independent proportionality factor –- intensity-related susceptibility~\cite{ZapKoz1, ZapKoz2}. As the response $R$ one may consider intensity of the transmitted, reflected, or scattered light, intensity of secondary emission, or even some non-optical quantities, like, say, amount of released heat. For the linear response of this kind, the superposition principle should be also satisfied (provided, of course, that the interference terms are averaged):
\begin{equation}
\label{lin:intensities}
R (I_1 + I_2) = R (I_1) + R(I_2),
\end{equation}
where $I_1$ and $I_2$ are the intensities of two incoherent fields.

Informative abilities of optical spectroscopy, however, are known to be immensely widened in nonlinear optics, where the relationship between $\bm P$ and $\bm E$ acquires a more complicated form~\cite{Bloembergen:432375,boyd2003nonlinear,shen1984principles}
\begin{equation}
\label{PEnonlin}
 P_i = \chi^{(1)}_{ij} E_j  + \chi^{(2)}_{ijk} E_jE_k + \chi^{(3)}_{ijkl} E_jE_kE_l + \ldots,   
\end{equation}
with larger number of invariants characterizing the medium and carrying additional information about the system. Here $\chi^{(1)}_{ij} \equiv \chi_{ij}$ is the linear susceptibility, $\chi^{(2)}_{ijk}$ and $\chi^{(3)}_{ijkl}$ are the second and third order nonlinear susceptibilities, and dots denote omitted higher order terms.
As a result, the nonlinear optical spectroscopy offers a great variety of novel  physical phenomena and fundamentally new methods of research  that make it possible to penetrate much deeper into the structure and symmetry of the medium, the nature and properties of the states responsible for the optical transitions, dynamics of optical excitations, etc.; see, e.g.,~\cite{boyd2003nonlinear,shen1984principles,Dantus2001coherent,Rottwitt2014nonlinear}. In particular, isotropic crystals, in nonlinear optics, may become optically anisotropic, forbidden transitions may become allowed, unresolved optical bands may reveal their inner structure, and so on.

Usually, the effects of nonlinear optics with their great informative abilities are observed at elevated light intensities, where the second- and higher order terms in Eq.~\eqref{PEnonlin} become important. The reverse statement also seems to be valid: when the light power density is low enough so that the terms $\propto E^2$, etc. are negligible, we remain in the realm of linear optics with its restricted abilities. This, however, is not the case for the SNS as we demonstrate below.

In this paper, we analyze and justify specific properties of the Faraday-rotation-based SNS, which, on the one hand, does not imply the use of strong optical fields, where the effects of higher-order susceptibilities become noticeable, but, on the other, exhibits properties that go far beyond the potentialities of conventional linear optics. These interesting properties have been discovered in the SNS studies during the last several years. We first overview specific features of the SNS that make it unique compared to other methods of linear optical spectroscopy, and then, on the basis of direct connection between the SNS and spin-flip Raman scattering, we present model consideration explaining these remarkable properties of the SNS.     

\section{Specific features of the spin noise spectroscopy}

\begin{figure}[t]
\includegraphics[width=\linewidth]{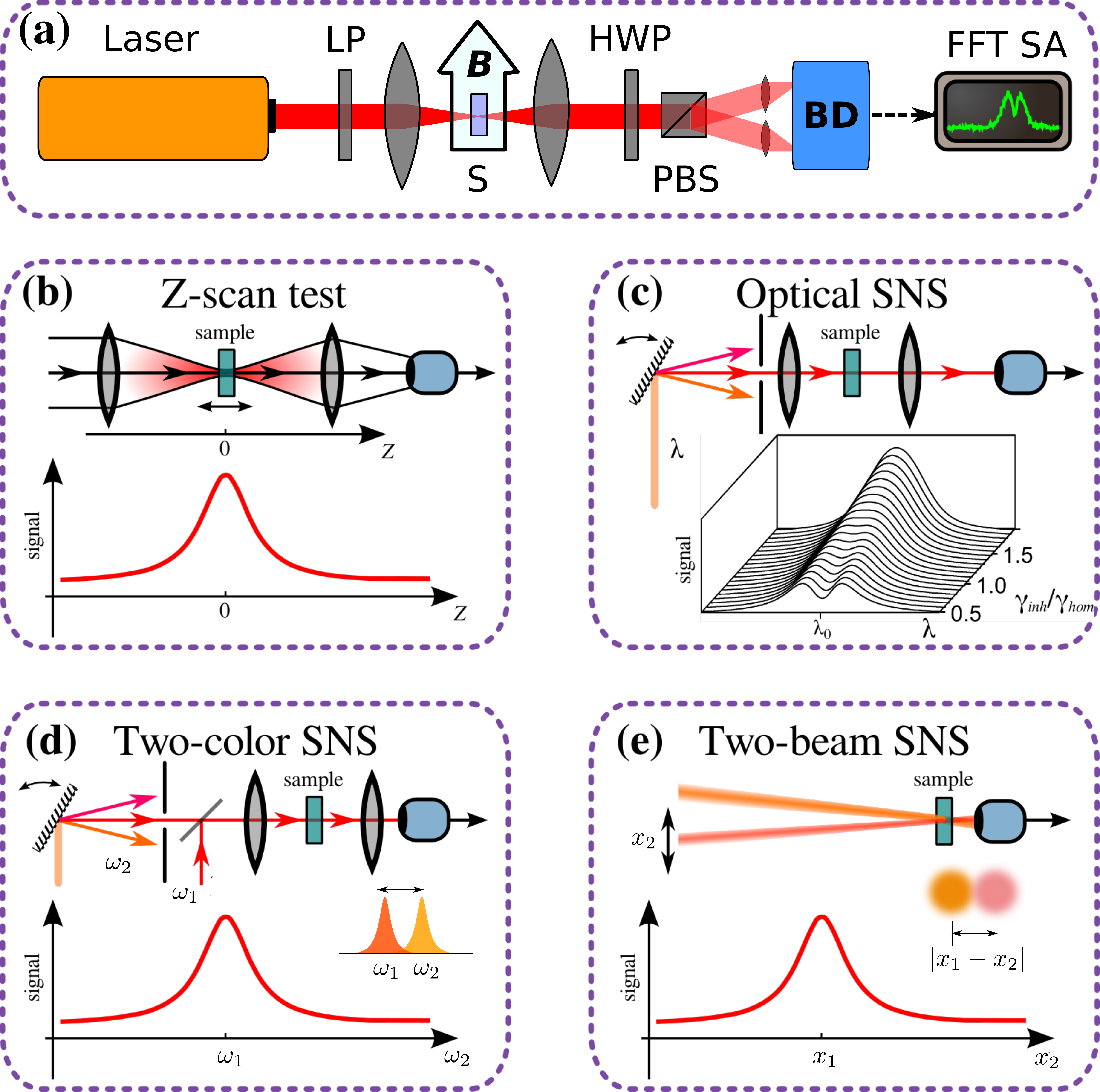}
\caption{{\textbf{Spin noise spectroscopy is a nonperturbative optical method of detecting magnetic resonance with abilities unusual for linear optics.} \textbf{(a)} Schematics of conventional SNS setup~\cite{Zapasskii:13}. LP is the linear polarizer, S is the sample, HWP is the half-wave plate, PBS is the polarization beam splitter, BD is the balanced detector, FFT SA is the fast-Fourier-transform spectrum analyzer; \textbf{(b)} $Z$-scan test and 3D tomography arise from an effective dependence of spin-noise signal on the light power density~\cite{Zscan2009}; \textbf{(c)} optical spectroscopy of spin noise allows penetration inside an inner structure of optical transitions~\cite{PhysRevLett.110.176601}; \textbf{(d)} two-color (``pump-probe'') spin noise spectroscopy allows measuring homogeneous linewidth of optical transitions~\cite{Yang:2014aa}; \textbf{(e)} two-beam (“pump-probe”) spin noise spectroscopy allows measuring spin transfer~\cite{twobeam}.}}\label{fig:0}
\end{figure}

We recall that the spin noise spectroscopy is a technique that allows one to detect magnetic resonance in the Faraday (or Kerr) rotation noise spectrum~\cite{aleksandrov81,Crooker_Noise,Oestreich:rev,Zapasskii:13}. Experimental arrangement commonly used for these measurements is shown in Fig.~\ref{fig:0}(a).  The probe light whose polarization noise is detected usually propagates across the applied magnetic field $\bm B$ its wavelength lies in the region of transparency of the sample so that the light does not perturb the system and does not change its properties. In this sense, the technique does not go beyond the bounds of linear optics. At the same time, it clearly shows properties that are usually characteristic of nonlinear optics only.

Specifically, the optical signal of spin noise shows positive reaction to the so-called $Z$-scan test intended to reveal optical nonlinearity~\cite{zscan1,zscan2}. As schematically shown in Fig.~\ref{fig:0}(b), the noise power of the sample drawn through the waist of a focused beam exhibits a peak centered at the point
of focus.  The fact that the mean square of the Faraday rotation noise is inversely proportional to the illuminated spot area~\cite{aleksandrov81,Mueller2010,gi2012noise} may be formally regarded as an evidence of dependence of the ``noise signal'' on the light power density,  typical for nonlinear optics. This property of SNS was used to propose a new method of three-dimensional tomography~\cite{Zscan2009}. 

Another interesting feature of the SNS is revealed when we consider it as a method of \emph{optical spectroscopy}, i.e. when the signal is studied  as a function of probe wavelength $\lambda$. It has been shown in Ref.~\cite{PhysRevLett.110.176601} that, due to sensitivity of the spin noise power to intrinsic structure of the optical spectrum, SNS allows one to resolve components of the spectrum unresolvable in conventional linear spectroscopy, to penetrate inside the profile of an inhomogeneously broadened line, and to determine its homogeneous width, Fig.~\ref{fig:0}(c).  

One more aspect that reveals unique properties of the SNS is related to the two-beam experiments that, evidently, do not have much sense in linear optics where different light beams do not interact anyway.  In particular, two probe beams with different frequencies $\omega_1$ and $\omega_2$ were used in \cite{Yang:2014aa} to monitor spin fluctuations in the inhomogeneous ensemble of quantum dots. Correlation between the Faraday-rotation noise in the two laser bheams arose only when the detuning $|\omega_1 - \omega_2|$ became comparable with (or smaller than) the homogeneous linewidth of the optical transition $\gamma$,~Fig.~\ref{fig:0}(d). This experiment made it possible to measure homogeneous width of a strongly inhomogeneously broadened band and thus to obtain information usually accessible only to methods of nonlinear optics, such as spectral hole burning, four-wave mixing, photon echo, etc.; see, e.g., \cite{holeburning,FWmixing,Meier-Koch}. 

The two-beam SNS-experiments with spatially separated beams make it possible, as was suggested in Ref.~\cite{Zapasskii:13}, to distinguish the spot illuminated by one of the beams with the aid of the other one, Fig.~\ref{fig:0}(e).  The cross correlation between the Faraday-rotation fluctuations of the two beams arises only if the beams overlap, because, in this case, their fluctuations are provided, at least partly, by the same spins. The situation may be different when the spins contributing to the detected noise are not localized. Then, the correlation between fluctuating signals of the two beams may persist (with a certain time delay) even when the light spots do not overlap. This experimental approach, as it was proposed in \cite{twobeam}, can be used to test spin transport under conditions of thermodynamic equilibrium. 

Thus, we see that the versions of SNS with two beams, in essence, realize a kind of \emph{pump-probe spectroscopy} without any optical nonlinearity, in spite of the fact that this illumination does not perturb the medium and, therefore, cannot be considered as a {\emph pump} in its conventional sense. Note that, in these experiments, the intensity superposition principle formulated by Eq. \eqref{lin:intensities} proves to be violated. 

These effects in SNS appear to be even more puzzling if we recall that shortly after experimental demonstration of spin noise, a deep relation between the SNS and coherent Raman scattering has been demonstrated by Gorbovitskii and Perel~\cite{gorb_perel}. It was shown that the effect of magnetic resonance in the Faraday rotation noise spectrum is a result of coherent Raman scattering of the probe beam in the forward direction (see below for details). In other words, the SNS is nothing but a modified spin-flip Raman spectroscopy, which differs from the latter by the method of measurement only: in Raman spectroscopy the scattered spectral component is detected in the straightforward way by means of optical spectroscopy, while in the SNS this is made using the heterodyning technique, so that the detected signal appears to be shifted from optical to radio frequencies. In terms of the Wiener-Khinchin  theorem, we may say that in Raman spectroscopy and in spin noise spectroscopy we study, respectively, correlation properties of the light field and light intensity. As a result, the Raman spectroscopy, in the noise-based modification, acquires new properties that are absent in the conventional optical spectroscopy of Raman scattering. So, it appears that this distinction in experimental techniques makes Raman spectroscopy and SNS methods drastically different.  To figure out the reason of this distinction, let us consider in more detail the process of signal formation in the SNS and in the conventional Raman spectroscopy. 

\section{Formation of the signals: Basic equations}

\begin{figure*}[t]
\includegraphics[width=\linewidth]{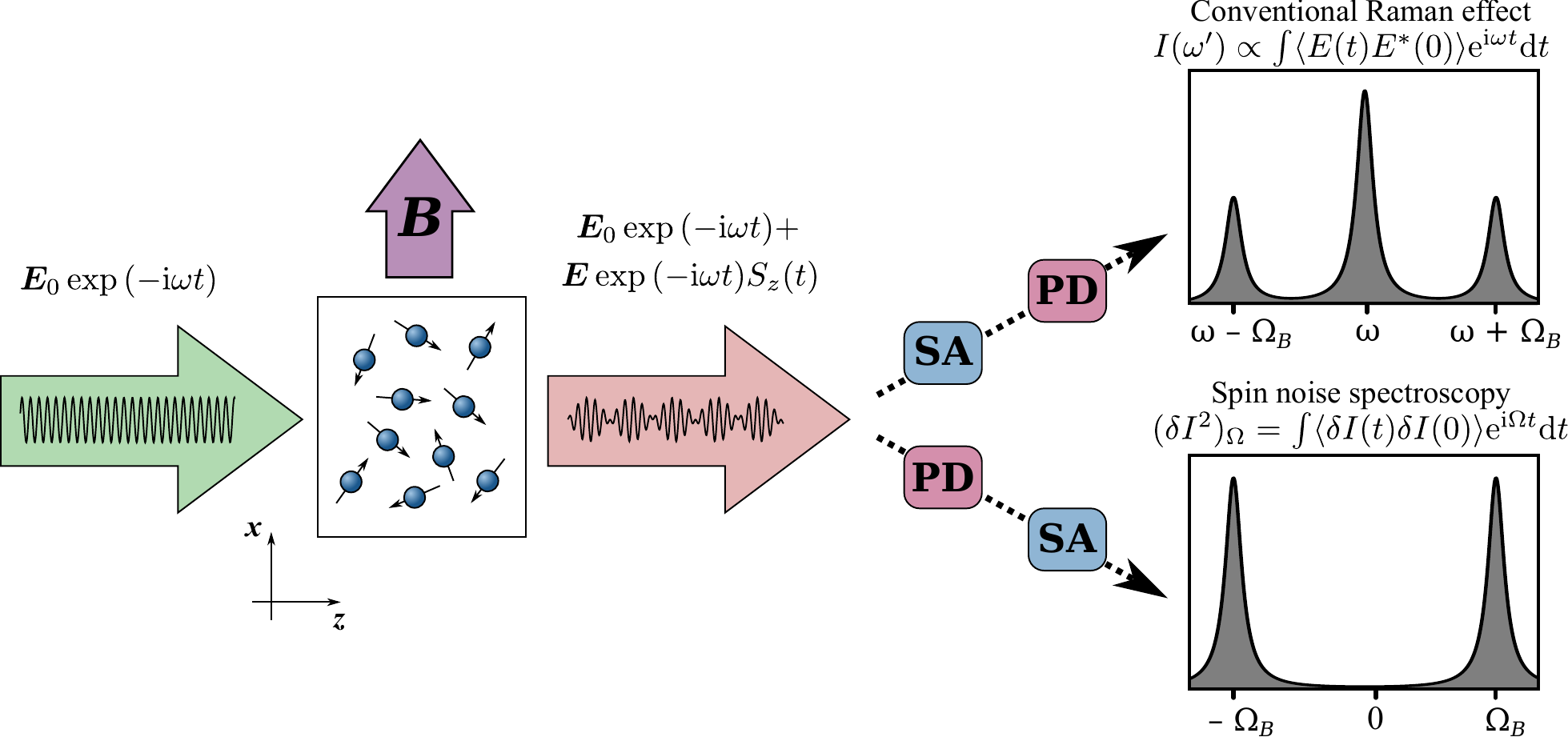}
\caption{
\textbf{Schematic representation of two experimental approaches to measuring spin-flip Raman scattering.} Transmitted field is modulated by spin fluctuations and, correspondingly, acquires Raman-shifted spectral sidebands. SA is the spectrum analyzer and PD is the photodetector. In conventional Raman scattering experiments, SA is the optical spectrometer, in spin noise spectroscopy experiments SA analyzes the light intensity spectrum in the range of radio-frequencies. }\label{fig:scatt}
\end{figure*}

Let us assume that the spin system under study is placed into an external magnetic field and irradiated by a monochromatic light beam propagating across the field. Consider the behavior of the Raman scattering signal related to light scattering by spin fluctuations $\delta \bm S$ in the media for the two above methods of the measurements: by direct detection of the Raman component in optical spectrum (in conventional Raman spectroscopy) and using the heterodyning method (in SNS). Schematically, the two possible detection schemes are shown in Fig.~\ref{fig:scatt}. It is noteworthy that the two approaches formally differ by the order of   arrangement of the spectrum analyzer and photodetector~\cite{Zapasskii:13,IntNoise}. In the conventional Raman spectroscopy, the scattered light is first passed through a spectrum analyzer (optical spectrometer) and then is converted into photocurrent (or photocharge). In the SNS, on the contrary, the scattered light is first converted into photocurrent, and only after that the spectral analysis of the photocurrent is carried out.

\begin{figure}[t]
\includegraphics[width=\linewidth]{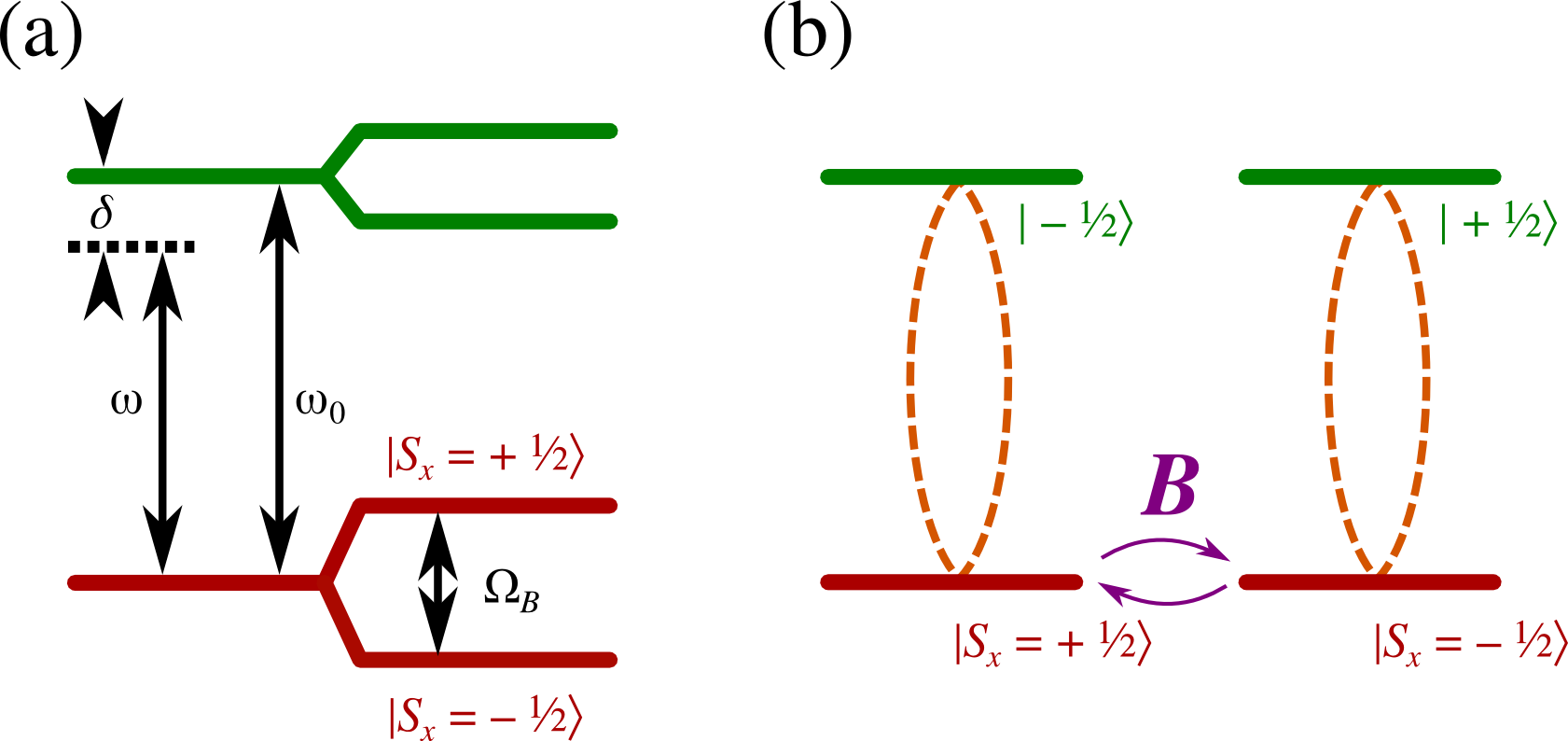}
\caption{
\textbf{Schematic energy diagram of the model four-level system under consideration.} \textbf{(a)} Eigenstates of the system $|S_x = \pm 1/2\rangle$  in the field $\bm B$ parallel to $x$-axis.  Double-headed arrows show the transition frequency, $\omega_0$, probe beam frequency, $\omega$, and Larmor frequency $\Omega_B=g\mu_B B/\hbar$, where $g$ is the Land\'{e} factor; $\delta$ is the detuning. \textbf{(b)} Schematics of the same diagram in the basis of nonstationary states $|S_z = \pm 1/2\rangle$ monitored by the light beam. Blue curved arrows depict mixing of the states by the transverse magnetic field. Dashed ellipses show the components of the polarization $\bm P$ of the two-level system.
}\label{fig:scheme}
\end{figure}

For the sake of an illustration we consider the simplest possible model of light interacting with a spin system: a four-level model, Fig.~\ref{fig:scheme}, where two ground states of the system are denoted as $S_z=\pm 1/2$ states and two excited states as $\pm 1/2$. Optical transitions between the ground and exited states are related to a dipole-allowed resonance responsible for the light scattering and spin noise. Such a model is widely used to describe light-matter interaction in atomic and semiconductor systems~\cite{Carm,ivchenko05a}.

We are interested in manifestation of the spin-$S_z$ fluctuations in optical response of the system probed by an incident beam with the frequency $\omega$ close to the resonance in the system.  We denote as $P_+(P_-)$ the dipole moments of the system in $\sigma^+(\sigma^-)$ polarization, their dynamics is given by, e.g.~\cite{glazov:review,gi2012noise}:
%
\begin{equation}
\label{dipoles}
\mathrm i \dot{P}_\pm = (\omega_0 - \mathrm i \gamma) P_\pm - (1/2 \mp S_z) \frac{|d|^2}{\hbar}  E_\pm \mathrm e^{-\mathrm i \omega t},
\end{equation}
where  $d$ is the dipole matrix element of the transition to the excited states, $\omega_0$ is the resonance frequency, $\omega$ and $\omega_0$ are assumed to be close to each other in order to neglect all other resonances in the system, $\gamma$ is the damping rate, and $S_z \equiv S_z(t)$ is the fluctuating spin $z$ component. In derivation of Eq.~\eqref{dipoles}, the incident field amplitude was assumed small enough, so that the probability to find the system in the excited states is negligible and spin fluctuations are quasi-static. Equation~\eqref{dipoles} can be solved with the result:
\begin{equation}
\label{Ppm}
P_\pm  = P_{\pm}^0(t) + P_{\pm}'(t),
\end{equation}
where   
\begin{equation}
P_{\pm}^0(t) = \frac{\mathrm i|d|^2E_\pm}{2\hbar}  \frac{\mathrm e^{-\mathrm i \omega t}}{\mathrm i(\omega_0 - \omega) + \gamma},
\end{equation}
oscillates at a frequency of the incident wave, while the second term 
\begin{equation}
\label{PfluctT}
P_{\pm}'(t) = -\frac{\mathrm i|d|^2E_\pm}{\hbar}  \int \frac{\mathrm d\Omega}{2\pi} \frac{\mathrm e^{-\mathrm i \omega t-\mathrm i \Omega t} \tilde S_z(\Omega)}{\mathrm i(\omega_0 - \omega - \Omega) + \gamma},
\end{equation}
where $\tilde S_z(\Omega) = \int \mathrm d t S_z(t)\exp{(\mathrm i \Omega t)}$,  is nonmonochromatic and related to spin fluctuations. Note, that both contributions to the dielectric polarization $P_\pm^0$ and $P_\pm'$ are linear in the incident radiation amplitude.

In what follows, we consider the forward and backward scattering geometries in co-circular polarizations ($\sigma^+$ to be specific); such geometries of Raman scattering are denoted as $z(\sigma^+\sigma^+)z$ and $z(\sigma^+\sigma^+)\bar z$, see \cite{ivchenko05a} for notations. To simplify the analysis, we assume also that the sample is illuminated by a plane-parallel beam of light with homogeneous intensity distribution. The amplitude of the detected electric field is given by\footnote{The electric field at the detector is, as a rule, inhomogeneous due to, e.g.,  inhomogeneity of the sample, effects of diffraction, etc. The way how the signal is picked up and/or averaged in the particular setup may change the results quantitatively.}
\begin{equation}
\label{forward}
E_{+}(t) = \alpha E_+ \mathrm e^{-\mathrm i \omega t} + \beta [P_+^0(t)+P_+'(t)],
\end{equation}
where the parameters $\alpha$ and $\beta$ depend on the light propagation geometry, properties of the spin ensemble, and details of the experimental setup~\cite{gi2012noise}. The detected field contains, along with the incident wave, the contribution oscillating at the same frequency, which corresponds to the resonance fluorescence, as well as the fluctuating contribution $\propto P_+'(t)$, which is responsible both for the Raman scattering and spin noise signals.

The above basic equations for the scattered electromagnetic field allow us to make clear the nature of distinction between the SNS and conventional Raman spectroscopy. Below, we consider in detail the two experimental approaches and demonstrate the origin of their different informative abilities.

\section{Discussion}

In the conventional spin-flip Raman spectroscopy, the recorded optical spectrum of the scattered light
\begin{equation}
\label{intensity}
I(\omega') \propto \Re \int_{-\infty}^\infty \langle E_+(t) E_+^*(0)\rangle \mathrm e^{\mathrm i \omega't} \mathrm dt.
\end{equation}
Note that  $I(\omega')$ gives the intensity of spectral component of electric field with the frequency~$\omega'$. In Eq.~\eqref{intensity}, the electric field is expressed via the fluctuating polarization $P'(t)$, Eq.~\eqref{PfluctT}, and the angular brackets denote averaging over the spin fluctuations in the corresponding quadratic combinations of the polarization as:
\begin{equation}
\label{spin:corr}
\langle \tilde S_z(\Omega_1) \tilde S_z^*(\Omega_2) \rangle = 2\pi\delta(\Omega_1 - \Omega_2)(S_z^2)_{\Omega_1},
\end{equation}
with $(S_z^2)_\omega$ being the spin noise power spectrum~\cite{gi2012noise}. Making use of Eqs.~\eqref{forward} -- \eqref{spin:corr}, we obtain the scattered intensity spectrum (apart from the $\delta$-function peaks arising from the incident field and Rayleigh scattering) in the form
\begin{multline}
\label{Raman:int}
I_{sc}(\omega') \propto |E_+|^2 (S_z^2)_{\Omega} \frac{\Gamma_0^2}{(\omega_0-\omega')^2+\gamma^2} \\
= |E_+|^2 (S_z^2)_{\Omega}\frac{\Gamma_0^2}{\delta^2+\gamma^2}.
\end{multline}
Here, $\Gamma_0 \propto |d|^2$ is the radiative decay rate of the excited state, $\Omega = \omega' - \omega$ is the Raman shift, and $\delta = \omega_0-\omega$ is the detuning between the optical resonance and the incident light frequency. Here and below, $\Omega$ in the denominators is neglected compared with $\gamma$ or $\delta$.

One can see that the shape of the Raman line follows the spin fluctuation spectrum multiplied by the Lorenzian profile of the optical response. In particular, the spin fluctuation spectrum in the transverse magnetic field is given by~\cite{gi2012noise}:
\begin{equation}
\label{sns:z}
(S_z^2)_\Omega = \frac{N}{4}\left(\frac{\tau_s}{1+(\Omega-\Omega_B)^2\tau_s^2} + \frac{\tau_s}{1+(\Omega+\Omega_B)^2\tau_s^2} \right),
\end{equation}
where $\tau_s$ is the spin relaxation time and $\Omega_B$ is the Larmor frequency.  Correspondingly, the Raman spectrum contains two peaks at $\omega' = \omega \pm \Omega_B$, see Fig.~\ref{fig:scatt}. It is important that the Raman scattering intensity scales linearly with the number of spins $N$ within the irradiated volume. Indeed, the total spin fluctuation of $N$ identical systems is $(S_z^2)_\Omega \propto N$, so, as follows from  Eq.~\eqref{Raman:int}, if we compress the light beam leaving its total power the same, the increase in the power density will be compensated by a decrease of the number of spins, with the total scattered intensity remaining the same. As a result, in particular, the forward scattering of a focused beam will be equally contributed by its cross-sections both in the vicinity of the focal point and far away from it.
Actually, this is exactly what should be expected for the effect of linear optics. It should be emphasized that the scattered intensity proves to be proportional to the spin fluctuation squared and, hence, does not depend on sign of the spin fluctuation~$S_z$. 

In the spin noise spectroscopy, by contrast, what we measure is the spectrum of the light intensity rather than intensity spectrum of the light field. In other words, SNS is actually a version of the {\emph light intensity spectroscopy} first proposed by Forrester and coworkers~\cite{Forrester} and then developed for spatially separated fields by Hanbury-Brown and Twiss \cite{Twiss}, see also \cite{Sillitto, Fellgett, Rebka}. 
Formally, we need to average the $|E(t')|^2$ over the time $T_a$ which, on the one hand, exceeds by far the period of electromagnetic field oscillations ($T_a \gg 2\pi/\omega$, $2\pi/\omega_0$) but, on the other, is much smaller than characteristic fluctuation times in the system. As a result,
\[
I_a(t) \propto \frac{1}{T_a} \int_{t-T_a/2}^{t+T_a/2} |E(t')|^2 \mathrm dt'.
\]
Such an averaged intensity contains a stationary (time-independent) contribution and a fluctuating one caused by interference of the incident wave, $E_+\exp{(-\mathrm i \omega t)}$, and the fluctuating field $\alpha P_+'(t)$~\cite{gorb_perel}:
\begin{equation}
\label{I:fluct}
\delta I(t) \propto |E_+|^2 \Gamma_0 \int \frac{\mathrm d\Omega}{2\pi} \Re\left[\frac{\mathrm e^{-\mathrm i \Omega t} \tilde S_z(\Omega)}{ (\omega_0 - \omega - \Omega) -\mathrm i \gamma} \right].
\end{equation}
 This contribution is present only in the forward- or backward-scattering geometry,\footnote{In the latter case the field propagating backwards contains the contribution $\propto E_+$ due to specular nonresonant reflection from the sample surface.} because in the case the scattering at an arbitrary angle, the electric field at the detector does not contain the incident wave at all, i.e. $E(t) \propto P_+^0(t)+P_+'(t)$.\footnote{There is also an interference of $P_+^0$ and $P_+'$, which is briefly discussed below.}

The signal again is proportional to the light intensity, but now it is linearly related to the spin fluctuation which, in turn, varies in a square-root way with the number of spins. 
It follows from Eq.~\eqref{I:fluct} that power spectrum of the intensity noise has the form:\footnote{One can arrive to Eqs.~\eqref{corr:I} and \eqref{corr:I:1} evaluating fourth-order correlator $\int \int \langle E_+^*(t) E_+^*(t+\tau) E_+(t+\tau) E_+(t)\rangle \exp{(\mathrm i \Omega \tau)} \mathrm dt \mathrm d\tau$.}
\begin{equation}
\label{corr:I}
(\delta I)^2_\Omega \propto |E_+|^4 (S_z^2)_\Omega \frac{\Gamma_0^2 \delta^2}{(\delta^2+\gamma^2)^2}.
\end{equation}
Equation~\eqref{corr:I} clearly demonstrates that the intensity fluctuations in the forward-scattering geometry are proportional to the electron spin noise spectrum.
Dependence of the intensity fluctuations on optical frequency (or detuning $\delta$) corresponds to the results of \cite{gi2012noise} for the Faraday rotation fluctuations. Moreover, at zero detuning, $\delta=0$, $(\delta I)^2_\Omega$ vanishes.  
It is worth noting that the interference of $P_+^0$ and $P_+'$ contributions to the field in forward direction yields ellipticity fluctuations. Such a contribution has the form
\begin{multline}
\label{I:fluct1}
\delta I'(t) \propto |E_+|^4 \Gamma_0^2 \times\\
 \int \frac{\mathrm d\Omega}{2\pi} \Re\left\{ \frac{\mathrm e^{-\mathrm i \Omega t} \tilde S_z(\Omega)}{[(\omega_0 - \omega - \Omega) -\mathrm i \gamma][(\omega_0-\omega)+\mathrm i \gamma]} \right\},
\end{multline}
and the corresponding intensity noise power spectrum, in agreement with \cite{gi2012noise}, can be presented as
\begin{equation}
\label{corr:I:1}
(\delta I')^2_\Omega \propto |E_+|^4 (S_z^2)_\Omega \frac{\Gamma_0^4 }{(\delta^2+\gamma^2)^2}.
\end{equation}
This contribution, unlike the one given by Eq.~\eqref{corr:I}, does not vanish at $\delta=0$. We note that in actual SNS experiments one uses linearly polarized probe beam and monitors fluctuations of the Faraday/Kerr rotation or ellipticity, which makes it possible to distinguish contributions of Eq.~\eqref{corr:I} and \eqref{corr:I:1}, Ref.~\cite{gi2012noise}.

Thus, the detected intensity noise spectrum, Eq.~\eqref{corr:I} or \eqref{corr:I:1}, reflects the spectrum of spin noise and, as sketched in Fig.~\ref{fig:scatt}, has two peaks at the frequencies $\Omega =\pm \Omega_B$ corresponding to spin precession in the transverse field.

It also follows from Eqs.~\eqref{corr:I} and \eqref{corr:I:1} that the detected signal in the SNS is proportional to the fourth power of electric field and to the spin noise power. Now, if we compress the light beam keeping the same its total power, the signal will not remain the same, but rather will grow in inverse proportion with the beam area. As a result, now the forward scattering of the focused beam will be mainly contributed by the spins in its focal region. This makes it possible to probe the medium by drawing the waist of a tightly focused light beam ($Z$-scan) and thus to realize the spin-noise-based 3D-tomography proposed in~\cite{Zscan2009}. 

Another highly important feature of the detected signal in this geometry is that, in contrast to the signal of conventional Raman scattering, the fluctuating  
contribution to the scattered intensity, Eq.~\eqref{I:fluct}, is sensitive to the {\it sign} of spin fluctuation $S_z$. It means that it 
 may interfere (constructively or destructively) with another spin noise signal. Particularly, for two incident beams with optical frequencies $\omega_1$ and $\omega_2$ the intensity fluctuations $\delta I_1(t)$ and $\delta I_2(t)$ in accordance with Eq.~\eqref{I:fluct} are proportional to ($\Omega \ll \gamma$, $|\omega_0 - \omega_1$, $\omega_0-\omega_2|$)
\[
\int \frac{\mathrm d\Omega}{2\pi} \Re\left[\frac{\mathrm e^{-\mathrm i \Omega t} \tilde S_{1,z}(\Omega)}{ (\omega_0 - \omega_1) -\mathrm i \gamma} \right]
\]
{and}
\[ \int \frac{\mathrm d\Omega}{2\pi} \Re\left[ \frac{\mathrm e^{-\mathrm i \Omega t} \tilde S_{2,z}(\Omega)}{ (\omega_0 - \omega_2) -\mathrm i \gamma} \right],
\]
respectively, where $\tilde S_{1,z}(\Omega)$, $\tilde S_{2,z}(\Omega)$ are the Fourier components of the spin fluctuations in the volumes illuminated by the corresponding beam. The power spectrum of cross correlation noise of the two beams is given by
\begin{equation}
\label{cross}
(\delta I_1 \delta I_2)_\Omega \propto \frac{(\tilde S_{1,z}\tilde S_{2,z})_\Omega}{[(\omega_0-\omega_1)^2+\gamma^2][\omega_0-\omega_2)^2+\gamma^2]}.
\end{equation}
Clearly, if the beams do not overlap, the cross correlation or interference noise is absent since spin fluctuations $\tilde S_{1,z}(\Omega)$ and  $\tilde S_{2,z}(\Omega)$ correspond to different ensembles.\footnote{There could be minor contributions if spins can freely propagate and go between the spots illuminated by the first and the second beams, c.f. Ref.~\cite{twobeam}.} If two beams overlap in space, the fluctuations become correlated, since $\tilde S_{1,z}(\Omega)$ and $\tilde S_{2,z}(\Omega)$ contain the contributions from the same spins. Moreover, the interference contribution~\eqref{cross} demonstrates strong spectral sensitivity: It is greatly enhanced if $\omega_1$ and $\omega_2$ are close to each other and differ by less than $\gamma$ from the resonance frequency $\omega_0$.\footnote{In inhomogeneous systems Eq.~\eqref{cross} should be modified following approaches of Refs.~\cite{gi2012noise,PhysRevLett.110.176601,Yang:2014aa} to take the inhomogeneity into account.} As a matter of fact, the presence of interference contribution, Eq.~\eqref{cross}, violates intensity superposition principle, Eq.~\eqref{superposition}, even though two optical beams come from different sources and optical coherence between them is absent.

 Due to this additional degree of freedom, the SNS can be used to reveal spatial or spectral coherence in spin fluctuations. 
Thus, in these light-intensity noise measurements, we lose the coherence in optics, but acquire spin coherence that provides us additional degrees of freedom absent in conventional spectroscopy of linear response. In particular, possibility of penetrating into the inhomogeneously broadened line results from spin-related interference of different spectral components of the probe beam. Positive response to the $Z$-test results from spin interference of different spatial components of the light beam. The same reason underlies the mentioned in \cite{Zapasskii:13}  possibility to detect illuminated spot be means of SNS.

\section{Conclusion}

The aim of this paper is to attract attention to the spin noise spectroscopy and to justify its remarkable informative capabilities, which may seem, at first glance, paradoxical: On the one hand, the SNS is considered to be just a version of spin-flip Raman spectroscopy and does not imply any optical nonlinearity of the medium, while, on the other, it offers abilities unusual for linear optics. In particular, the SNS signal proves to be dependent on the light power density, it allows one to resolve inner structure of inhomogeneously broadened spectra, and to realize pump-probe spectroscopy with no optical nonlinearity. We show that these unique features of the SNS result, in fact, from linear relationship between the signal and spin fluctuation (in contrast to quadratic one for Raman spectroscopy), Eq.~\eqref{I:fluct}. This makes SNS sensitive to spatial and spectral correlations of the polarization noise and thus provides additional informative abilities inaccessible for conventional Raman spectroscopy and even for linear optics in general. 

These specific properties of the spin noise spectroscopy are determined, in essence, by correlation nature of the light-intensity measurements, which provides the intensity signal with a fluctuating part having an effective sign. As a result, the two light-intensity signals may combine constructively or destructively exactly like it occurs with regular sign-alternating signals. Actually, a similar reason underlies specific properties of the conventional light-intensity-noise spectroscopy~\cite{IntNoise}, which is known to be capable of getting very special information about the source of the field \cite{Jeltes,Bromberg, Lahini, Perrin}.  In the latter case, however, the light field under study is usually produced by a primary source, and information extracted from the correlation measurements is not related to inner structure of the self-luminous medium. 

It is noteworthy that similar advantages of the heterodyning technique should be revealed in other effects of inelastic light scattering and may be of interest for a wide range of applications.

\par\bigskip

\textbf{Acknowledgements}
We are grateful to E.L. Ivchenko, K.V. Kavokin, and A.V. Kavokin for discussions and I.I. Ryzhov for technical help. This work was supported by RFBR, RF
President grant NSh-1085.2014.2, Russian Ministry
of Education and Science (Contract No. 11.G34.31.0067 with SPbSU and leading
scientist A.V. Kavokin), SPbSU grant 11.38.277.2014 and EU project SPANGL4Q.

%



\end{document}